\documentclass[10pt,journal,compsoc]{IEEEtran}
\ifCLASSOPTIONcompsoc
\usepackage[nocompress]{cite}
\else
\usepackage{cite}
\fi
\ifCLASSINFOpdf
\usepackage[pdftex]{graphicx}
\graphicspath{{figures/}}
\else
\fi
\usepackage{amsmath}
\usepackage{balance}       % to better equalize the last page
\usepackage{graphics}      % for EPS, load graphicx instead 
\usepackage[T1]{fontenc}   % for umlauts and other diaeresis
\usepackage{txfonts}
\usepackage{mathptmx}
\usepackage[pdflang={en-US},pdftex]{hyperref}
\usepackage{color}
\usepackage{booktabs}
\usepackage{textcomp}

\usepackage{placeins}

\usepackage{microtype}        % Improved Tracking and Kerning
\usepackage{ccicons}          % Cite your images correctly!
\newcommand{\etal}{et~al.}
\newcommand{\eg}{e.g., }
\newcommand{\ie}{i.e., }
\newcommand{\pvalue}{$p$-value}
\newcommand{\pvalues}{$p$-values}
\newcommand{\logit}{\textrm{logit}}

\begin{document}
	%Copyright notice
	%\pagestyle{empty}

	%
	% paper title
	% Titles are generally capitalized except for words such as a, an, and, as,
	% at, but, by, for, in, nor, of, on, or, the, to and up, which are usually
	% not capitalized unless they are the first or last word of the title.
	% Linebreaks \\ can be used within to get better formatting as desired.
	% Do not put math or special symbols in the title.
	\title{Can visualization alleviate dichotomous thinking? Effects of visual representations on the cliff effect}

	%
	%
	% author names and IEEE memberships
	% note positions of commas and nonbreaking spaces ( ~ ) LaTeX will not break
	% a structure at a ~ so this keeps an author's name from being broken across
	% two lines.
	% use \thanks{} to gain access to the first footnote area
	% a separate \thanks must be used for each paragraph as LaTeX2e's \thanks
	% was not built to handle multiple paragraphs
	%
	%
	%\IEEEcompsocitemizethanks is a special \thanks that produces the bulleted
	% lists the Computer Society journals use for "first footnote" author
	% affiliations. Use \IEEEcompsocthanksitem which works much like \item
	% for each affiliation group. When not in compsoc mode,
	% \IEEEcompsocitemizethanks becomes like \thanks and
	% \IEEEcompsocthanksitem becomes a line break with idention. This
	% facilitates dual compilation, although admittedly the differences in the
	% desired content of \author between the different types of papers makes a
	% one-size-fits-all approach a daunting prospect. For instance, compsoc 
	% journal papers have the author affiliations above the "Manuscript
	% received ..."  text while in non-compsoc journals this is reversed. Sigh.
	\author{Jouni~Helske,~ Satu~Helske,~
		Matthew~Cooper,~
		Anders~Ynnerman,~\IEEEmembership{Member,~IEEE},~
		and~Lonni~Besan\c{c}on % <-this % stops a space
		\IEEEcompsocitemizethanks{
			\IEEEcompsocthanksitem J. Helske was with Department of Science and Technology, Link\"oping University, Campus
			Norrk\"oping, SE-602 74 Norrk\"oping, Sweden, and now with Department of Mathematics and Statistics, University of Jyv\"askyl\"a, FI-40014 Jyv\"askyl\"a, Finland.\protect\\
			% note need leading \protect in front of \\ to get a newline within \thanks as
			% \\ is fragile and will error, could use \hfil\break instead.
			E-mail: jouni.helske@iki.fi
			\IEEEcompsocthanksitem S. Helske is with Department of Social Research, University of Turku, FI-20014 Turku, Finland.
			\IEEEcompsocthanksitem M. Cooper, A. Ynnerman, and L. Besan\c{c}on are with Department of Science and Technology, Link\"oping University, Campus
			Norrk\"oping, SE-60274 Norrk\"oping, Sweden.}% <-this % stops an unwanted space
		\thanks{Published in the IEEE Transactions on Visualization and Computer Graphics. DOI: \url{10.1109/TVCG.2021.3073466}}}

	\IEEEtitleabstractindextext{%
		\begin{abstract}
			Common reporting styles for statistical results in scientific articles, such as \pvalues\ and confidence intervals (CI), have been reported to be prone to dichotomous interpretations, especially with respect to the null hypothesis significance testing framework. For example when the \pvalue\ is small enough or the CIs of the mean effects of a studied drug and a placebo are not overlapping, scientists tend to claim significant differences while often disregarding the magnitudes and absolute differences in the effect sizes.
			This type of reasoning has been shown to be potentially harmful to science.
			Techniques relying on the visual estimation of the strength of evidence have been recommended to reduce such dichotomous interpretations but their effectiveness has also been challenged. We ran two experiments on researchers with expertise in statistical analysis to compare several alternative representations of confidence intervals and used Bayesian multilevel models to estimate the effects of the representation styles on differences in researchers' subjective confidence in the results. We also asked the respondents' opinions and preferences in representation styles. Our results suggest that adding visual information to classic CI representation can decrease the tendency towards dichotomous interpretations -- measured as the `cliff effect': the sudden drop in confidence around \pvalue\ 0.05 -- compared with classic CI visualization and textual representation of the CI with \pvalues. All data and analyses are publicly available at \href{https://github.com/helske/statvis}{https://github.com/helske/statvis}.
		\end{abstract}
		
		% Note that keywords are not normally used for peerreview papers.
		}

	% make the title area
	\maketitle

	% To allow for easy dual compilation without having to reenter the
	% abstract/keywords data, the \IEEEtitleabstractindextext text will
	% not be used in maketitle, but will appear (i.e., to be "transported")
	% here as \IEEEdisplaynontitleabstractindextext when the compsoc 
	% or transmag modes are not selected <OR> if conference mode is selected 
	% - because all conference papers position the abstract like regular
	% papers do.
	\IEEEdisplaynontitleabstractindextext
	% \IEEEdisplaynontitleabstractindextext has no effect when using
	% compsoc or transmag under a non-conference mode.

	% For peer review papers, you can put extra information on the cover
	% page as needed:
	% \ifCLASSOPTIONpeerreview
	% \begin{center} \bfseries EDICS Category: 3-BBND \end{center}
	% \fi
	%
	% For peerreview papers, this IEEEtran command inserts a page break and
	% creates the second title. It will be ignored for other modes.
	\IEEEpeerreviewmaketitle
	
	\IEEEraisesectionheading{\section{Introduction}\label{sec:introduction}}
	
	One of the most common research questions in many scientific fields is ``Does X have an effect on Y?'', where, for example, X is a new drug, and Y a disease.  Often the question is reduced to ``Does the average effect of X differ from zero?'', or ``Does X significantly differ from Z?''. There are various statistical approaches available for answering this question, and many ways to report the results from such analyses. In many fields, null hypothesis significance testing (NHST) has long been the de-facto standard approach. NHST is based on the idea of postulating a ``no-effect'' null hypothesis (H0) which the experimenter aims to reject. An appropriate test statistic, based on assumptions about the data and model, is then calculated together with the corresponding \pvalue, the probability of observing a result at least as extreme as the one observed under the assumption that H0 is true. Small \pvalues\ indicate incompatibility of the data with the null model,  assuming that the assumptions used in calculating the \pvalue\ hold.
	
	The ongoing `replication crisis'~\cite{Pashler2012}, especially in social and life sciences, has produced many critical comments against arbitrary \pvalue\ thresholds and significance testing in general (\eg~\cite{Amrhein2019,Wasserstein2019, McCloskey2008}). As a solution to avoid so-called dichotomous thinking -- strong tendency to divide results into significant or non-significant -- some are even arguing for a complete ban on NHST and \pvalues. Such a policy has also been adopted by some journals: \eg in 2015, the Journal of Basic and Applied Social Psychology banned both \pvalues\ and confidence intervals (CIs)~\cite{Trafimow2015}, and more recently the Journal of Political Analysis banned the use of \pvalues~\cite{gill2018}. 
	
	Despite the critique, significance testing is likely to remain a part of a scientist's toolbox. Because many of the problems with NHST are due to misunderstandings among those who conduct statistical analyses as well as among those who interpret results, work has also been conducted in making it easier to avoid common pitfalls of NHST either by altering the way analyses are conducted \cite{kay2016researcher, kruschke2018bayesian, statrethinkingbook} or how the results are presented \cite{Cumming:2012:NS, Calin-Jageman:2018:NSB, correll-gleicher-2014, Kalinowski:2018:STCI, Dragicevic:2019:EMA}. 
	Instead of arguing for better methodological solutions, such as Bayesian approaches, here we study whether different styles of visual representation of common statistical problems could help to alleviate dichotomous thinking which can be approximated by studying the so-called \emph{cliff effect} \cite{Lai:2010:DT}. The cliff effect is a term used for the large difference in how the results are interpreted despite only small numerical differences in the estimate and $p$-value \cite{Rosenthal:1963:ILS} (\eg the estimated effect of 0.1 with a corresponding $p$-value of 0.055 may be deemed not significant while an effect of~0.11 with a $p$-value of 0.045 may be claimed to be significant). 
	
	In this paper we focus on the effect of visualization styles on confidence profiles (the perceived confidence--p-value relationship) and in particular on the magnitude of the cliff effect. To study the potential cliff effect of various representation styles for statistical results, we conducted two experiments on researchers who are experienced in using and interpreting statistical analyses. We showed participants results from artificial experiments using different representation styles and asked the respondents how confident they were in that the results showed a positive effect (experiment~1) or a difference between two groups (experiment~2). We also asked the respondents to give comments on the different styles and to rank them according to their personal preference. We analysed the answers from the experiments using Bayesian multilevel models. These results are easy to interpret and at the same time allow us to avoid the problems we aimed at studying (i.e. dichotomous thinking and the cliff effect). 
	
	Three earlier studies somewhat resemble our experimental setting. First, we use and compare similar visualizations of the uncertainty of the sample mean as Correll and Gleicher \cite{correll-gleicher-2014}. We, however, focus on a different research question and correspondingly a different target population. Correll and Gleicher were interested in the communication of mean and error to a general audience while our interest is in dichotomous thinking, by measuring the confidence profile and the cliff effect.  We are interested in the interpretations of quantitative scientific results, which requires a fundamental understanding of statistics. Hence we target researchers whose dichotomous interpretations can have adverse effects on conclusions and gained knowledge. Furthermore, addressing the question left open in \cite{correll-gleicher-2014}, we also collected qualitative data on researchers' preferences for different visualization styles.
	Lai \cite{Lai:2010:DT}, on the other hand, similarly to us focused on the magnitude of the cliff effect and the shape of the confidence profile. He manually categorized respondents' confidence profiles into four different categories, discarding a large proportion of answers which did not fit into any category. Instead of comparing different visualization methods, he only used classic CI visualization.
	Third, Belia \etal~\cite{Belia:2005:RMC} had a similar approach in that they also focused on experts' (mis)conceptions of confidence intervals (with researchers in medicine and psychology). Their focus was not on dichotomous thinking but on finding the positioning of two confidence intervals so that the respective groups would be deemed ``statistically significantly different''. Similarly to Lai, they only used classic CI visualization. 
	Our contributions are as follows: 1)~This is the first study to examine the effects of visualization styles of CIs on confidence profiles and dichotomous thinking (using the cliff effect as a proxy) among researchers. 2)~We are the first to study researchers' preferences on novel visualization styles in this context. 3)~We introduce the use of flexible and easy-to-interpret Bayesian framework for the analysis of confidence profiles and representation preferences. 4)~As a contribution to open science, codes for the online experiment, data and all analyses are publicly available and fully reproducible.
	Our results suggest that despite the increased debate around NHST and related concepts, the problem of dichotomous thinking persists in the scientific community, but that certain visualization styles can help to reduce the cliff effect and should be used and studied further.
	
	\section{Background and Related Work}\label{sec:background}
	
	In this paper, our main focus lies in whether and how different visualizations can help in reducing the cliff effect  among researchers making interpretations of inferential statistics. 
	We first briefly present the basic definition and interpretation of the confidence interval (CI), which is a common choice for assessing the uncertainty of a point estimate (e.g., sample mean) and has sometimes been suggested to reduce dichotomous interpretations. We then discuss the problem of dichotomous thinking in scientific reporting before presenting related literature and the visual representations used in our experiments.
	
	\subsection{Confidence Interval for Sample Mean}
	
	Given a sample of values $x_1,\ldots,x_n$ from a normal distribution with unknown mean $\mu$ and variance $\sigma^2$, the 95\% confidence interval for the mean is computed using a sample mean $\bar x$, sample standard deviation $s$, sample size $n$ and $t$-distribution:
	\begin{equation}
	\bar x \pm t_{\alpha/2}(n - 1) \frac{s}{\sqrt{n}},
	\end{equation}
	where $t_{\alpha/2}(n - 1)$ is the critical value from $t$-distribution with $n - 1$ degrees of freedom and significance level $\alpha$ (typically $0.05$). The interpretation of the (95\%) CI is somewhat complicated: Given multiple 95\% CIs computed from independent samples, on average 95\% of these intervals will contain the true expected value $\mu$. It is important to note that, 
	given a single sample and the corresponding CI, we cannot infer whether the true population mean, $\mu$, is contained within the CI or not \cite{Neyman} although it has a direct connection to NHST in that the 95\% CI represents the range of values of $\mu$ for which the difference between $\mu$ and $\bar x$ is not statistically significant at the 5\% level.
	
	\subsection{The Problem of Dichotomous Thinking in Science}
	\label{sec:dichotomous}
	
	Let us suppose that through an experiment we obtain a \pvalue\ of $p=0.048$. Most researchers would consider this strong enough evidence against H0. If, however, we obtained $p$-value of 0.058 many researchers, despite the small difference, would follow the recommendations of  colleagues and textbooks, consider this as not enough evidence against H0 \cite{gigerenzer2004mindless}. This type of reasoning, often called \emph{dichotomous thinking} or \emph{dichomotous inference} has been shown to be potentially harmful to science \cite{amrhein2018inferential,Amrhein2019,Blakeley:2017:SSDE,Cockburn:2020:TRC,Rafi2020}. It has been said to be one of the reasons for the replication crisis \cite{swikatkowski2017replicability,amrhein2018inferential,Cockburn:2020:TRC} or to lead to ``absurd replication failures [with] compatible results''~\cite{Amrhein2019}. While dichotomous thinking has been heavily criticized by scholars (\eg \cite{amrhein2018inferential,Blakeley:2017:SSDE,Cumming:2012:NS,Dragicevic:2014:RAH,Dragicevic:2016:FSC}, it seems to be 
	persistent in many fields including HCI~\cite{Besancon:2019:CPDI} and empirical computer science~\cite{Cockburn:2020:TRC}.

	In 2016, the confusion, misuse and critique around \pvalues\ led the American Statistical Association (ASA) to issue a statement on \pvalues\ and statistical significance. ASA stated that proper inference must be based on full and transparent reporting and computing, and that a single number (\pvalue) is not equal to scientific reasoning. Many other authors have criticized the whole NHST approach due to increased dichotomous thinking based on arbitrary thresholds~\cite{amrhein2018inferential,Amrhein2019, Dragicevic:2016:FSC, gelman2017threshold, gigerenzer2004mindless, gigerenzer2018statistical}, common misinterpretations of \pvalues\ (\eg the fallacy of accepting H0 \cite{Altman:1995:AOE}, reading \pvalues\ as the probability that H0 is true), as well as the several questionable research practices that often come with the use of NHST including $p$-hacking (testing a number of hypotheses until a low \pvalue\ is found), HARKing (presenting a post-hoc hypothesis as an a priori hypothesis), selective outcome reporting, and the file-drawer effect (limiting publication to only statistically significant results) \cite{Cockburn:2020:TRC,Norbert1998, John:2012:MPQRP,Ioannidis:2005:WMP,Simmons:2011:FPP,Ulrich:2017:SPP,Wicherts:2016:DF}. Additionally, sometimes \pvalues\ are reported without effect sizes, although a \pvalue\ itself does not help readers determine the practical importance of the presented results. It should be noted that it is likely that many of these issues relating to the data-led analysis (see the ``garden of forking paths''\cite{Gelman2013} ) are typically not intentional, and can occur in a broader scope than just NHST.
	
	Due to all of the issues around \pvalues, some researchers have recommended either to replace or complement them with CIs~\cite{Besancon:2017:SD,Cumming:2012:NS,Cumming:2014:NS,Dragicevic:2016:FSC}. The argument is that CIs could reduce dichotomous interpretations as they represent both the effect size and the sampling variation around this value. CIs, however, are also prone to misinterpretation, simply because their interpretation is 
	not very intuitive~\cite{Hoekstra:2014:RMCI, Belia:2005:RMC}. 
	CIs have also been reported to lead to dichotomous thinking 
	\cite{Besancon:2019:CPDI,Lai:2010:DT, Hoekstra:2012:CIMD}.
	
	The term cliff effect was coined by Rosenthal and Gaito in their study \cite{Rosenthal:1963:ILS} on 19 researchers in psychology. Their findings were later replicated by Nelson \etal \cite{Nelson:1986:ISS} on a larger sample (85 psychologists). Poitevineau and Lecoutre~\cite{Poitevineau:2001:CEO} showed that only a small fraction of their participants adopted a dichotomous all-or-none strategy, while Lai \cite{Lai:2010:DT} showed that milder tendencies to dichotomous thinking exist.
	Another study by Poitevineau and Lecoutre~\cite{Lecoutre:2003:ESI} suggested that even statisticians were not immune to misinterpretations and dichotomous thinking. However, due to the previous focus on restricted populations (mainly psychologists) and also because some of the details of the experiments have not been fully presented (such as the exact question asked to the participants), it is difficult to assess whether these findings would hold in a more general population of researchers.
	
	Previous studies on interpretation of \pvalues\ and CIs have suggested that there are two to four confidence interpretation profiles~\cite{Poitevineau:2001:CEO, Lai:2010:DT,Kalinowski:2018:STCI}. While some individual variation and hybrid interpretation styles are likely to exist, due to historical reasons it is likely that the main profiles are the all-or-none category (related to Neyman-Pearson significance testing) and the gradually decreasing confidence category (related to Fisher's significance testing approach). See, for example,~\cite{Perezgonzalez2015} for descriptions of the original approaches to significance testing by Fisher, and Neyman and Pearson as well as their connection to current NHST practice.
	
	Bayesian paradigm and replacing CIs with \emph{credible intervals} have been suggested as a solution to the problems with CIs and \pvalues \cite{kay2016researcher, kruschke2018bayesian,Scheibehenne:2016:BES,Wagenmakers:2007:PSP}.
	Compared to the CI, the credible interval has a more intuitive interpretation: given the model and the prior distribution of the parameter (\eg mean), the 95\% credible interval contains the unknown parameter with 95\% probability. Or perhaps even better, one can present the whole posterior distribution of the parameter of interest. Despite the benefits of the Bayesian approach,
	\pvalues\ and CIs are likely to remain in use in many scientific fields, despite their flaws. Hence it is of general interest to study whether the problems relating to dichotomous thinking can be alleviated by changing their typical representation styles.

	\subsection{Visualization of Uncertainty and Statistical Results}
	\label{sec:pastvis}
	
	Several visualization techniques have been designed to show the uncertainty of the estimation, with several advantages over the communication of a sole point estimate~\cite{Jung:2015:DUI,Wunderlich:2017:VDU}. Showing the theoretical or empirical probability distribution of the variable of interest is a commonly used technique.
	For example, \emph{probability density plots} are often used for describing the known distributions such as Gaussian distribution or estimated density functions based on samples of interest (\eg observed data or samples from posterior distributions in a Bayesian setting). \emph{Violin plots} \cite{Hintze1998} (also called \emph{eyeball plots} in~\cite{Spiegelhalter1999}) are rotated and mirrored kernel density plots, so that the uncertainty is encoded as the width of the `violin' shape. \emph{Raindrop plots}~\cite{Barrowman:2003:RP} are similar to violin plots but are based on log-density. The \emph{gradient plot} uses opacity instead of shape to convey the uncertainty (\eg~\cite{correll-gleicher-2014}), while \emph{quantile dot plots}~\cite{Fernandes:2018:UDU,Kay:2016:MBU} are discrete analogs of the probability density plot. 
	Various alternative representation styles specifically for CIs are commonly used (see, for example,~\cite{Cumming2007}). In order to remedy the misunderstanding and misinterpretation of CIs, Kalinowski \etal~\cite{Kalinowski:2018:STCI} designed the \emph{cat's eye confidence interval} which uses normal distributions to depict the relative likelihood of values within the CI (based on the Fisherian interpretation of the CI). 
	A violin plot with additional credible interval ranges are also used to depict arbitrary shaped (univariate) posterior distributions based on posterior samples, for example in the \texttt{tidybayes} R package (coined as the \emph{eye plot})~\cite{tidybayes}. Kale \etal~\cite{Kale:2019:HOP,Hofman:2020:VIU} studied animated \emph{hypothetical outcome plots} for interactive dissemination of statistical results. Going even further, Dragicevic \etal~\cite{Dragicevic:2019:EMA} propose the use of interactive explorable statistical analyses in research documents to increase their transparency. For a systematic review of uncertainty
	visualization practices, see Hullman \etal~\cite{Hullman:2019:POE}.
	
	Some  past  studies have focused  on comparing  different visual  representations of statistical results. Tak \etal~\cite{Tak:2014:PVU} examined seven different visual representations of uncertainty on 140 non-experts. Correll and Gleicher~\cite{correll-gleicher-2014} studied four different visualization styles for mean and error in several settings. Kalinowski \etal~\cite{Kalinowski:2018:STCI} compared students' intuitions when interpreting classic CI plots and cat's eye plots. Finally, the recent study by Hofman \etal~\cite{Hofman:2020:VIU} focused on the impact of presenting inferential uncertainty in comparison to presenting outcome uncertainty, and investigated the effect of different visual representations of effect sizes. With the exception of~\cite{Kalinowski:2018:STCI}, these studies have focused on testing lay-people, a population which can be expected to differ from researchers who have been trained to interpret $p$-values and CIs in their work.
	
	\section{Research Questions}\label{sec:questions}
	
	Taking inspiration from some of the approaches listed in Section~\ref{sec:background}, our work aims to evaluate the presence and magnitude of the cliff effect in textual and visual representation styles among researchers trained in statistical analysis. Our main goals were to investigate
	\begin{itemize}
		\item whether the cliff effect can be reduced by using different visual representations  instead of textual information and
		\item how researchers' opinions on, and preferences between, different representation styles differ.
	\end{itemize}
	
	More specifically, we were interested in whether the previously documented cliff effect in scientific reporting is reduced when the textual representation with explicit \pvalue\ is replaced with a traditional visualization of CI, and whether more complex visualization styles for the CI reduce the cliff effect. Regarding the former question, in line with previous research~\cite{Lai:2010:DT, Hoekstra:2012:CIMD}, we expected to find that CIs would not reduce the cliff effect, whereas regarding the latter question our hypothesis was that more complex visualization styles could reduce the cliff effect.  
	
	As our interest was in scientific reporting, we limited our sample to researchers with an understanding and use of statistics unlikely to be present with lay-people, and focused on static visualizations applicable in traditional scientific publications. While researchers are more familiar with confidence intervals and other statistical concepts, experts' interpretations can still exhibit various implicit biases and errors due to field's conventions and obtained education (see, e.g, \cite{McCurdy2018}). However, instead of studying the differences in various subgroups of scientific community, our interest is more about an "average researcher".
	
	\section{One-sample Experiment}
	\label{sec:experiment1}
	
\begin{figure*}
		\centering
		\includegraphics[width=2\columnwidth]{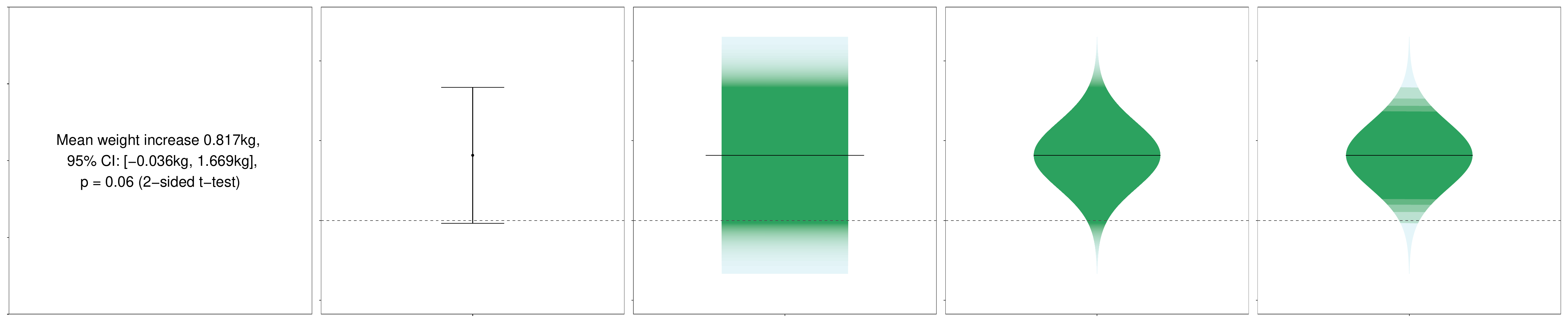}
		\caption{Representation styles used in the experiments: Textual version with \pvalue, classic 95\% confidence interval (CI), gradient CI plot, violin CI plot, and discrete violin CI plot.}
		\label{fig:teaser}
	\end{figure*}

	In the first experiment we are interested in potential differences in the interpretation of results of an artificial experiment when participants are presented with textual information of the experiment in a form of a \pvalue\ and a CI, a classic CI plot, a gradient CI plot, or a violin CI plot (see \autoref{fig:teaser} and the descriptions in \autoref{sec:conditions1}). The setting is simple yet common: we have a sample of independent observations from some underlying population, and we wish to infer whether the unknown population mean differs from zero. 
	
	\subsection{Conditions}
	\label{sec:conditions1}
	
	\subsubsection{Textual Information with $p$-value}
	
	Our first representation is text consisting of the exact \pvalue\ of a two-sided $t$-test, sample mean estimate and lower and upper limits of the 95\% CI (see the leftmost box of \autoref{fig:teaser} for the participant's view). This style is concise, contains information about the effect size and the corresponding variation (width of the CI), while the \pvalue\ provides evidence in the hypothesis testing style. While this format provides information on the effect size and uncertainty together with the \pvalue it can be argued that, due to the strong tradition in NHST, the inclusion of a \pvalue\ can cause dichotomous thinking even when accompanying CI information is provided. While the sample size is not stated in this format, that information was provided separately in our experiment for each condition as a part of the explanatory text.
	
	\subsubsection{Classic Confidence Interval Visualization}
	
	Confidence intervals and sample means are commonly visualized as line segments with end points augmented with horizontal lines (see \autoref{fig:teaser}). Compared with textual information, visual representation could be better at conveying the uncertainty. While the width of the horizontal lines of the CI does not have semantic meaning, it is sometimes argued (although we have found no studies to suggest this) that their width emphasises the limits of the CI and increases dichotomous inference, and intervals without the horizontal lines should be preferred. We chose the more traditional design (see the second box from left in Fig. \ref{fig:teaser}) as it is still commonly used and is a default in many statistical analysis packages such as SPSS.
	
	\subsubsection{Gradient Color Plot for CI (Gradient CI Plot)}
	
	In order to reduce the dichotomous nature of the classic CI visualization, we test the effect of using multiple overlaid confidence intervals with varying coverage levels and opacity. This is fairly common when presenting prediction intervals for future observations~\cite{hyndman2018}, but less so in the case of CIs. While using only a few overlaid CIs (\eg 80\%, 90\% and 95\%) is a more common practice, we decided to replicate the gradient plot format used in previous approaches~\cite{correll-gleicher-2014} which provides more emphasis on the 95\% interval and thus is more comparable with the classic CI approach. Our gradient CI plot contains a colored area of 95\% CI complemented with gradually colored areas corresponding to 95.1\% to 99.9\% CIs (with 0.1 percentage point increments), overlaid with a horizontal line corresponding to the sample mean (see the middle box in Fig.~\ref{fig:teaser}). The coloring was from hex color \texttt{\#2ca25f} to \texttt{\#e5f5f9} taken from ColorBrewer's 3-class BuGn palette~\cite{colorbrewer}. This format provides additional information, but gradual color changes can be difficult to interpret accurately, and from a technical point of view this format is also harder to create than classic CIs.
	
	\subsubsection{CI as $t$-violin Plot (Violin CI Plot)}
	
	While the gradient CI plot gives information about the uncertainty beyond the 95\% CI, we claim that the use of the rectangular regions with constant widths can be misleading. Therefore, as our fourth representation format (inspired by~\cite{correll-gleicher-2014, Kalinowski:2018:STCI}) we combine the gradient CI plot and the density of the $t$-distribution used in constructing the CIs (see the second box from right in Fig. \ref{fig:teaser}). More specifically, in the violin CI plot the shape corresponds to the case of computing a sequence of confidence intervals with very fine increments, with the width of each CI computed using the underlying $t$-distribution. The width of the violin at point $y$ is 
	\begin{align}
	p(\frac{\sqrt{n}(y - \bar x)}{s})\frac{\sqrt{n}}{s},
	\end{align} 
	where $p$ is the probability density function of the $t$-distribution with $n - 1$ degrees of freedom, $\bar x$ is the sample mean, and $s$ is the standard deviation.
	
	In the second experiment we also consider a more discretized version of the violin CI plot with gradually colored areas corresponding to the 80\%, 85\%, 90\%, 95\% and 99.9\% CIs (see the rightmost box in Fig. \ref{fig:teaser}).
	
	Violin CI plots are more challenging to create, and the probability density function style can lead to erroneous probability interpretations for which CIs cannot provide answers. On the other hand, the additional visual clues due to the shape can help overcome the difficulty of interpreting gradient colors.
	
	\subsection{Participants and Apparatus}
	
	The experiment was run as an online survey. Its preregistration is available on \url{https://osf.io/v75ea/}. As the preregistration states, the number of participants was not decided in advance but, instead, we aimed for the maximum number of participants in a given time frame. The end date of the experiment was fixed to 11 March 2019 so the survey was open for 21 days before we started to analyse the data. As stated in \autoref{sec:questions}, our goal, contrary to most of the previous work, was to understand how researchers interpret statistical results and therefore we aimed at recruiting academics familiar with statistical analysis. To recruit participants across various scientific disciplines, we initially contacted potential participants via email in several fields (namely Human Computer Interaction, Visualization, Statistics, Psychology, and Analytical Sociology, using personal networks), and the survey was also shared openly using the authors' academic profiles on Twitter and suitable interest groups on Reddit, LinkedIn, and Facebook.
	
	The eligibility criteria were 1) You understand English; 2) You are at least 18 years old; 3) You have at least a basic understanding of hypothesis testing and confidence intervals; 4) You use statistical tools in your research projects; 5) You are not using a handheld device such as tablet or phone to fill out the survey.
	To evaluate the validity of our sample, we asked for background information including participants' age, scientific field, highest academic degree, length of research experience, and data analysis tools commonly used. The codes for the experiment are available in supplementary materials on \url{https://github.com/helske/statvis}.
	
\begin{figure}
		\includegraphics[width=\columnwidth]{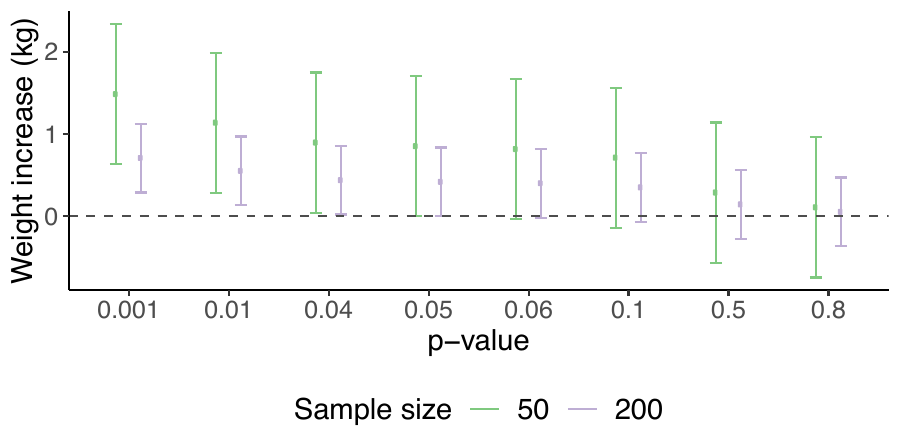}
		\caption{Configuration used in the one-sample experiment. See text for details.}
		\label{fig:configuration}
	\end{figure}

	There are multiple potential factors which could (although not necessarily should) have an effect on interpreting results of this simple experiment: \pvalue, total length of the confidence interval, effect size, sample size, and representation style. Since our focus was on the representation styles, and because we wanted to keep the survey short in order to increase the number of responses, we used a fixed set of \pvalues\ (0.001, 0.01, 0.04, 0.05, 0.06, 0.1, 0.5, 0.8), and a fixed standard deviation of 3. By defining also the sample size, the sample mean was then fully determined by these values. We used two sets of questions, one with a sample size of $n=50$ and another with $n=200$. Each participant saw the results corresponding to only one of these sets. \autoref{fig:configuration} shows the configurations as 95\% CIs with dots representing the means. The participants did not see the underlying $p$-values except in the textual representation style.
	
	During the experiment we displayed each trial to each participant (one at a time), and asked the following question: ``A random sample of 200 adults from Sweden were prescribed a new medication for one week. Based on the information on the screen, how confident are you that the medication has a positive effect on body weight (increase in body weight)?''. They answered on a continuous scale (100 points between 0 and 1, the numerical value was not shown) using a slider with labelled ends (``Zero confidence'', ``Full confidence''), which was explained to the participants as ``Leftmost position of the slider corresponds to the case ``I have zero confidence in claiming a positive effect,'' whereas the rightmost position of the slider corresponds to the case ``I am fully confident that there is a positive effect.'' The slider's `thumb' was hidden at first, in order to avoid any possible bias due to its initial position. It only became visible when the participant clicked on the slider. Finally, until the slider position was set participants could not proceed to the next question.
	
	Our small pilot study suggested that it was hard to understand the violin CI plot due to its non-standard meaning (participants were prone to misread the figure as a typical violin plot of empirical density of the data). Therefore, in order to explain the interpretation of the violin plot in this context, we had to also explain the basics of CI computations. To keep the complexity of all representations at the same level, we added explanatory texts to all conditions. We detail the impact of this decision in our discussions in \autoref{sec:discussion}
	
	In order to balance learning effects, the order of the four conditions (representation styles) was counterbalanced using Latin squares, and within each condition the ordering of trials was randomly permuted for each participant. At the end of the survey, participants had to give feedback on the representation formats and rank them from 1 (best) to 4 (worst). We gave participants the possibility to give equal rankings. They could also leave additional comments about the survey in general.
	
	We gathered answers from 114 participants, from which one participant was excluded because of nonsensical answers to the background questions. One of the background variables was an open-ended question about field of expertise. The answers included a range of disciplines that we categorized into four groups: ``Statistics and machine learning' (21 participants), ``VIS/HCI'' (34), ``Social sciences and humanities'' (32), and "Physical and life sciences" (26) (see supplementary material for more information).
	
	\subsection{Statistical methods}\label{sec:exp1-methods}
	
	All statistical analyses were done in the R environment~\cite{R} using the \texttt{brms} package~\cite{brms}. The visualizations of the results were created with the \texttt{ggplot2} package~\cite{ggplot2}. The collected data, scripts used for data analysis, additional analysis, and figures are available in supplementary material. We also created an accompanying R package \texttt{ggstudent}\footnote{\url{https://cran.r-project.org/package=ggstudent}} for drawing modified violin and gradient CI plots used in the study.
	
	To analyse the results we built a Bayesian multilevel model with participants' confidence as the response variable (values ranging from 0 to 1), and the underlying \pvalue\ and representation style as the main explanatory variables of interest.
	
	While we often perceive the probabilities and strength of evidence as having a linear relationship after logit-transformations of both variables \cite{Zhang2012}, in the case of significance testing with potential for dichotomous thinking this relationship is likely not true due to the potential cliff effect as well as the excess occurrence of low and high \pvalues\ indicating complete lack of evidence (0) or full confidence (1). Values 0 and 1 (15\% of all answers) are also problematic in the logit-transformation due their mapping to $\pm \infty$. Therefore, a simple linear model with logit-transformations of \pvalues\ and the confidence scores would not be suitable in this case.
	
	A typical choice for modelling proportions with disproportionately large numbers of zeros and ones 
	is the zero-one-inflated beta regression. However, as we wanted to incorporate the prior knowledge of the potential linear relationship of confidence and probabilities in the logit-logit-scale, instead of the zero-one-inflated beta distribution we created a piecewise logit-normal model\footnote{The distribution was changed from the preregistration as suggested by a reviewer and \cite{Zhang2012}.} with the probability density function (pdf) defined as
	\begin{equation}
	p(x)=\begin{cases}
	\alpha (1 - \gamma), & \text{if $x = 0$},\\
	\alpha \gamma, & \text{if $x = 1$},\\
	(1 - \alpha) \phi(\logit(x), \mu, \sigma), & \text{otherwise}.\\
	\end{cases}
	\end{equation}
	Here $\alpha = P(x \in \{0, 1\})$ is the probability of answering one of the extreme values (not at all confident or fully confident), whereas $\gamma = P(x = 1 \mid x \in \{0, 1\})$, is the conditional probability of full confidence given that the answer is one of the extremes\footnote{While generating data from this distribution is straightforward, the expected value of this distribution is analytically intractable. However, this can be easily computed via Monte Carlo simulation.}. Thus these two parameters model the extreme probability of answers, and when the answer is between the extremes, we model it with the logit-normal distribution ($\phi(x)$ is the pdf of the normal distribution parameterized with mean $\mu$ and standard deviation $\sigma$). Explanatory variables can be added to the model to predict $\alpha$, $\gamma$, $\mu$, and $\sigma$, using the log-link for $\sigma$, the logit-link for $\alpha$ and $\gamma$, and the identity-link for $\mu$. In comparison to the frequentist approach, such as standard generalized linear (mixed) models or analysing only simple descriptive statistics, our Bayesian model allows us to take into account the uncertainty of the parameter estimation and more flexible model structures. We can also make various simple probabilistic statements based on the posterior distributions of this model such as the probability that the cliff effect is higher with $p$-values than with classic CI. For further information about Bayesian modelling in general see, for example,~\cite{bda3}.
	
	\subsection{Results}
	
	\subsubsection{Confidence Profiles and Cliff Effects}
	
	As the first step, we checked some descriptive statistics of the potential cliff effect, defined as
	\[
	\delta = \textrm{E}\left[\textrm{confidence}(p=0.04) - \textrm{confidence}(p=0.06)\right],
	\]
	\ie the average difference in confidence between cases $p=0.04$ and $p=0.06$. Table~\ref{tab:desc1} shows how gradient and violin CI plots have a somewhat smaller drop in confidence when moving from $p=0.04$ to $p=0.06$ compared to the textual representation and the classic CI visualization.
	
	% Thu Jul 11 13:22:47 2019
	\begin{table}[t]
		\centering
		\caption{The sample mean, standard deviation, standard error of the mean, and the 2.5th and 97.5th percentiles of the difference in confidence when $p=0.04$ and $p=0.06$ in the first experiment.}
		\begin{tabular}{lrrrrr}
			\hline
			& Mean & SD & SE & 2.5\% & 97.5\% \\ 
			\hline
			Textual & 0.19 & 0.27 & 0.03 & -0.19 & 0.72  \\ 
			Classic CI & 0.23 & 0.25 & 0.02 & -0.05 & 0.84 \\ 
			Gradient CI & 0.10 & 0.24 & 0.02 & -0.37 & 0.74\\ 
			Violin CI & 0.13 & 0.20 & 0.02 & -0.16 & 0.62\\ 
			\hline
		\end{tabular}\label{tab:desc1}
	\end{table}
	
	To analyse the data and the potential cliff effect in more detail, we used the Bayesian multilevel model described in \autoref{sec:exp1-methods}. Due to the setup of the experiment, participants' answers were influenced by the information on the screen, which in turn depended on the underlying \pvalue, visualization style, and sample size. Sample size itself should not have an effect on the answers, which was indeed confirmed by preliminary analysis (see supplementary material), so we dropped that variable from further analysis. Due to the potential cliff effect we wanted to allow different slopes of the confidence curve for the cases when $p<0.05$ and $p>0.05$. With regards to the case of $p=0.05$ we allowed an extra drop in confidence via an indicator variable $I(p=0.05)$, as it was not clear whether this boundary case should be on the ``significant'' or ``not significant'' side (i.e., whether the cliff effect was due to the drop just before or after 0.05). Regarding the probability of an extreme answer, the relationship with respect to the \pvalue\ was assumed to be non-linear so we treated the \pvalues\ as a categorical variable. For the conditional probability of full confidence $\gamma$ we used the \pvalue\ as a categorical variable with a monotonic effect (using the simplex parameterization suggested in \cite{burkner_charpentier}), but grouped $p>0.05$ values together.
	
	As it was reasonable to assume that participants used different scales of confidence in their answers (e.g., some participants were always very confident), we included individual-level random intercepts for $\mu$, $\alpha$ and $\sigma$. We also allowed the effects of visualization and the underlying $p$-value to vary between participants by including corresponding random coefficients in the model. We ran various posterior predictive checks~\cite{Gabry2019} to assess that the model fits the data reasonably well (see the supplementary material). The final model structure, written using the extended Wilkinson-Rogers syntax \cite{Wilkinson1973, nlme} was chosen as follows:
	
	\begin{align}
	\begin{split}
	\mu        &\sim viz \cdot I(p < 0.05) \cdot \logit(p) + viz \cdot I(p = 0.05) \\ 
	& + (viz + I(p < 0.05) \cdot \logit(p) +  I(p = 0.05) \mid id),\\
	\alpha     &\sim  p \cdot viz + (1 \mid id),\\
	\gamma     &\sim mo(p),\\
	\sigma     &\sim viz + (1 \mid id),
	\end{split}
	\end{align}
	where $p$ is a categorical variable defining the true \pvalue, logit($p$) is a continuous variable of the logit-transformed \pvalue, $mo(p)$ denotes a monotonic effect of the \pvalue, the dot corresponds to interaction (\ie $I(p = 0.05) \cdot viz$ implies both the main and two-way interaction terms) and $(z \mid id)$ denotes participant-level random effect for variable $z$.
	
	Given this model, in a presence of a cliff effect we should observe a discontinuity in an otherwise linear relationship between the true \pvalue and reported confidence (when examined in the logit-logit scale). We used the relatively uninformative priors: N$(0,5)$ regression coefficients, N$(0,3)$ for the intercept terms, and half-N$(0, 2)$ for all standard deviation parameters, LKJ(1) prior~\cite{LEWANDOWSKI20091989} for the correlation matrices of random effects, and symmetric Dirichlet(1) prior for the coefficients of the monotonic effect.
	
	Consistent with the Bayesian paradigm, we chose this model over simpler submodels (where some of the interactions or random effects are omitted)~\cite{kruschke2014}. This model integrates over the uncertainty regarding the model parameters, with coefficient zero corresponding to a simpler model where the term is omitted from the model. However, as a sensitivity check, we also estimated several submodels of this model. These gave very similar results, so the reported results were insensitive to specific model choice.
	
\begin{figure}[t]
		\centering
		\includegraphics[width=\columnwidth]{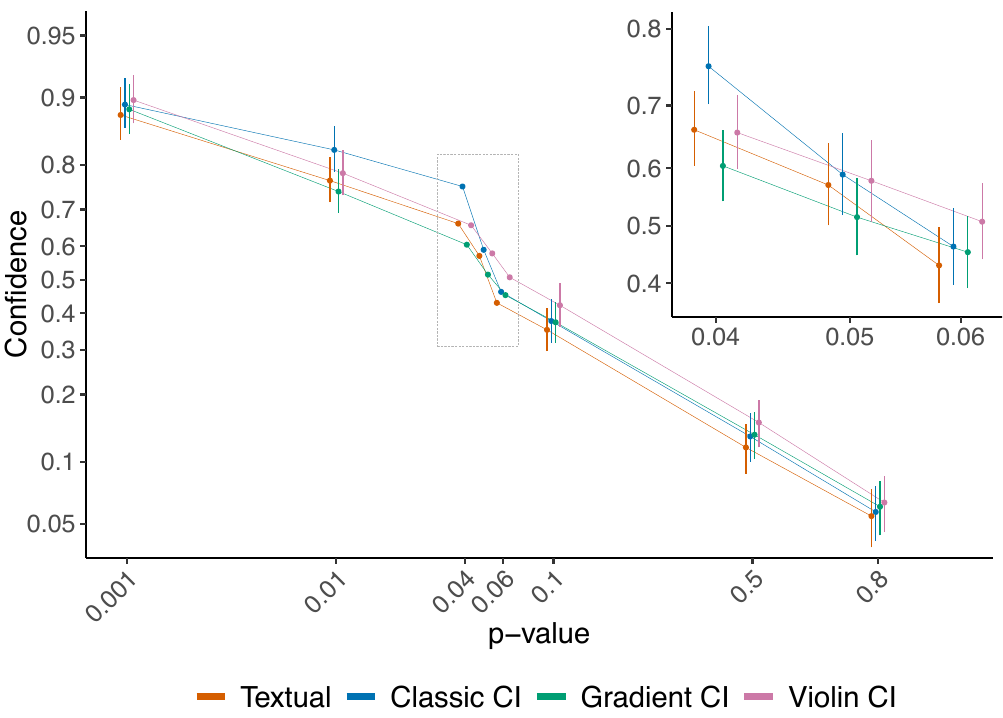}
		\caption{Posterior means of confidence and corresponding 95\% credible intervals for different visualization styles in the first experiment, on the logit-logit-scale. Here, a discontinuity in an otherwise linear relationship indicates a cliff effect. The zoom-in plot shows the uncertainty of the estimates when $0.04 \leq p \leq 0.06$.}
		\label{fig:pcurve1}
	\end{figure}

	\autoref{fig:pcurve1} shows the posterior mean curves of confidence (vertical lines corresponding to the 95\% credible intervals\footnote{For readers new to the credible interval, we refer to section~\ref{sec:dichotomous}.}) with respect to the underlying true \pvalues\ used to generate the data. These are based on the population level effects: the expected confidence of an average participant (an individual whose random effects are 0).
	
	We observe at least some kind of a cliff effect -- a sudden drop in confidence -- with all representation styles. Within the ``statistically significant region'' (\ie when $p<0.05$) the slope of the confidence level in relation to the underlying \pvalue\ is the least steep for the classic CI visualization, but there is a large drop in confidence when moving to $p>0.05$, even larger than with the textual information. The textual representation with $p$-value, on the other hand, behaves similarly to the violin CI plot until $p=0.05$, after which the confidence in the \pvalue\ representation drops below all other techniques. The gradient CI plot and the violin CI plot both have a smaller -- although visible -- drop in confidence and also otherwise show a similar pattern, except that the confidence level of the gradient CI plot is constantly below that of the violin CI plot. The range of confidence is very similar across all representation styles, which suggests that the smaller cliff effect of gradient CI and violin CI plot is not to due to overall smaller confidence (not an unreasonable assumption given their more fuzzy nature compared to classic CI). There were no clear differences in the probabilities of an extreme answer (``zero confidence'' or ``full confidence'') between the visualization styles (see the supplementary material). 
	
	\autoref{fig:cliff1} shows the posterior distributions of the drop in confidence, $\delta$, for different visualizations. These show that the drop is the largest with classic CI and the smallest (and nearly identical) with gradient and violin CI visualizations. Textual representations with $p$-values position between these (somewhat closer to the classic CI). The magnitude of the drop in the classic CI (mean of 0.29) is close to a third of the range of the confidence scale and twice as large as the drop in the gradient and violin CIs (means of 0.15). While there is some overlap between these distributions, when comparing the pairwise posterior probabilities that the $\delta$ of one visualization style is greater than that of an alternative style for an average participant (\autoref{tab:postprob1}), we see clear differences between the styles: Classic CI leads to larger drop than textual $p$-values, and both of these lead to larger drops than Gradient CI and Violin CI (all these comparisons have probabilities close to 1). Note that, unlike the interpretation of \pvalues, the numbers in \autoref{tab:postprob1} are actual probabilities that the average drop in confidence around $p=0.05$ is larger with one style than the other.
	
	\begin{figure}[t]
		\centering
		\includegraphics[width=\columnwidth]{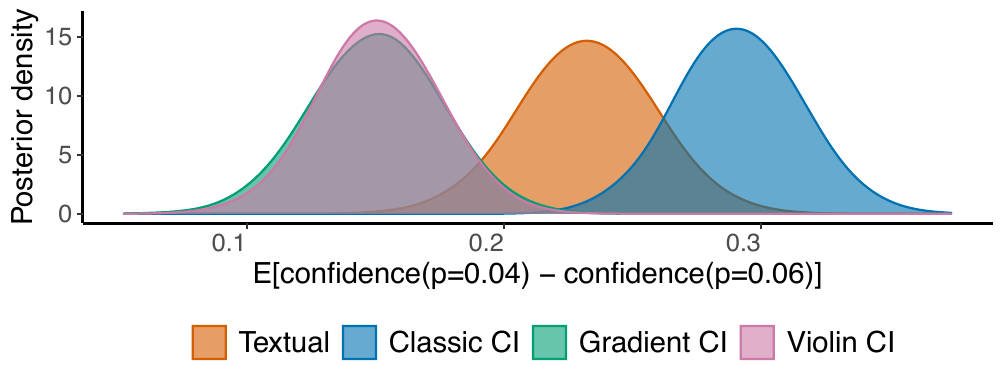}
		\caption{Posterior distributions of $\delta$, the drop in confidence around $p=0.05$, for different representation styles in the first experiment. Note that the distributions of the gradient CI and the violin CI on the left-hand side are almost completely overlapping.}
		\label{fig:cliff1}
	\end{figure}
	
	\begin{table}[t]
		\centering
		\caption{Posterior probability that $\delta$, the drop in confidence around $p=0.05$, is larger for representation style on the row than the representation style on the column.}
		\begin{tabular}{lrrrr}
			\hline
			& Textual  & Classic CI & Gradient CI & Violin CI \\ 
			\hline
			Textual     & -  & 0.01 & 1.00   & 1.00  \\ 
			Classic CI  & 0.99 & -  & 1.00   & 1.00 \\ 
			Gradient CI & 0.00 & 0.00 & -    & 0.49 \\ 
			Violin CI   & 0.00 & 0.00 & 0.51 & - \\ 
			\hline
		\end{tabular}\label{tab:postprob1}
	\end{table}
	
	As a secondary analysis, we also estimated a model with categorized expertise value as a predictor (with interactions with visualization and \pvalue). When averaging (i.e. marginalizing) over the expertise, the results were similar to the main model. The expertise-specific examinations, however, revealed some differences between the groups. Most notably we observed the largest cliff effects in the Stats/ML group (for all representation styles), while in the Phys/Life group there were only small differences in the confidence profiles by representation style. When comparing the magnitudes of $\delta$, the ordering of the representation styles was the same across all expertise groups (as seen in the main results). Due to space restrictions see the supplementary material for more detailed results.
	
	\subsubsection{Subjective Rankings}
	
\begin{figure}[t]
		\includegraphics[width=\columnwidth]{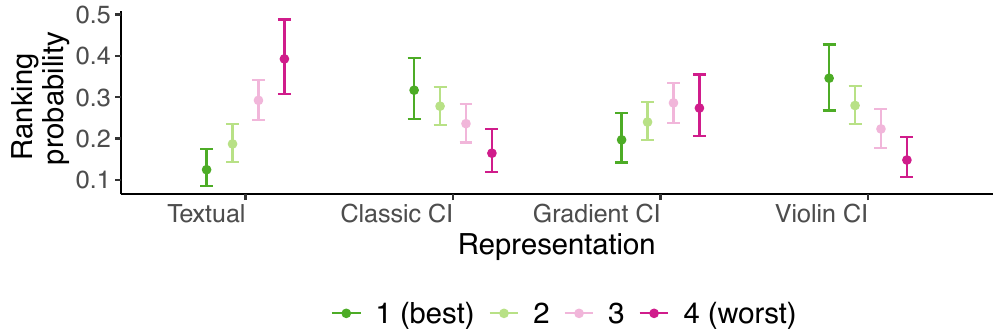}
		\caption{Subjective ranking probabilities and the corresponding 95\% credible intervals for visualization styles of the first one-sample experiment. A higher value for rank 1 indicates preference for the method while a higher value for rank 4 indicates distaste.}
		\label{fig:rankings}
	\end{figure}

	To analyse the subjective rankings of the representations, we estimated a Bayesian ordinal regression model where we used visualization style to predict the observed rankings (with participant-level random intercept). \autoref{fig:rankings} shows the results from this model as a probability that the visualization style obtains a certain rank. We see that \pvalue\ typically obtains the worst rank (4), while violin CI and classic CI are the most preferred options with approximately equal probabilities for ranks 1 and 2. Gradient CI seems to divide opinions, with close to equal probabilities for each rank.
	
	\subsubsection{Qualitative Feedback}\label{sec:subjective1}
	
	At the end of the experiment, participants were invited to comment on the limitations and benefits of each technique. The fully categorized and raw data is available in supplementary materials,
	but we summarize the main points here. The following summaries were created by one of the authors before seeing any of the other results. Concerning \pvalues, participants reported them to be easy to read and accurate ($\times$ 40 participants). However, participants also stated that they could hinder the readability of a paper if many of them had to be reported ($\times$ 11), that they could be difficult to interpret ($\times$ 33), that some expertise was needed to understand them ($\times$ 10), and that text-only might make readers focus on \pvalues\ exclusively ($\times$ 7). Furthermore, some participants explained that a visualization would have made the analysis much easier, in particular for the confidence interval ($\times$ 22).
	The condition with classic confidence intervals was said to be a standard ($\times$ 19) that allows quick analysis with clear figures ($\times$ 42) and that scales very well to multiple comparison ($\times$ 11). However, participants also reported that this visual representation was missing information ---likelihood of the tails for instance--- and that it should be augmented with more statistical information ($\times$ 33). Additionally, they were said to possibly foster dichotomization ($\times$ 10).
	Violin CI plots were judged to be visually pleasing ($\times$ 8), to provide all the statistical information that classic confidence intervals fail to provide ($\times$ 31) and to help avoiding the dichotomization pitfall ($\times$ 5). Nonetheless, some participants stated that they were representing too much information ($\times$ 4), that they might require training as they are not often used ($\times$ 17), and that the gradient at the tails was hard to see ($\times$ 13). In addition to this, some participants explained that such plots could be misunderstood  due to their similarity with the violin plot ($\times$ 6).
	Finally, the gradient CI plots were reported to be visually pleasing ($\times$ 5), to provide more information than a classic confidence interval ($\times$ 20), to help avoiding dichotomization ($\times$ 6). In addition to this, participants stated (either as a positive or negative point) that the cut off after 95\% was difficult to assess visually ($\times$ 9) which could also help reduce dichotomized interpretations. Participants also noted that the gradient was hard to distinguish ($\times$ 9), that making inferences based on gradient plots could be more difficult ($\times$ 11) and that the width was unnecessary visual information because it does not encode anything ($\times$ 13).
	
	\section{Two-sample Experiment}
	\label{sec:exp2}
	
\begin{figure}[t]
		\includegraphics[width=\columnwidth]{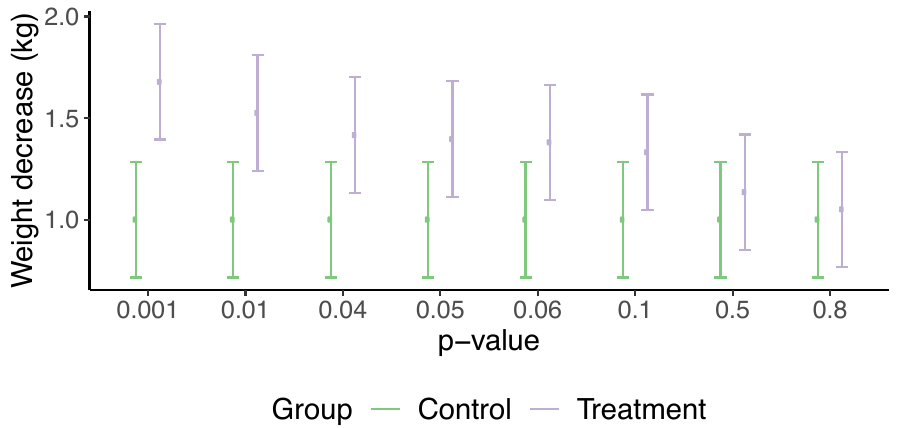}
		\caption{Configuration used in the second experiment.
		}
		\label{fig:configuration2}
	\end{figure}

	After conducting the first experiment, we deployed a second survey with a similar framing, but this time instead of comparing the base value of zero, the task was to compare means of independent ``treatment'' and ``control'' groups, as in~\cite{Belia:2005:RMC}. While it is often recommended that instead of comparing intervals of two (potentially dependent) samples it is better to compare intervals of the difference~\cite{cumming2005}, nevertheless these types of multiple interval visualizations are commonly seen in scientific publications. Similar to our first controlled experiment, this study was also preregistered\footnote{\url{https://osf.io/brjzx/?view_only=e481a9ad345e4e689799d65d988c1c5f}}, with supplementary material available at Github.\footnote{https://github.com/helske/statvis} \autoref{fig:configuration2} shows the configuration used in this second experiment. 
	
	\subsection{Conditions, Participants and Apparatus}
	
	The conditions and overall design of the study were the same as the one-sample experiment except that the textual \pvalue\ representation was replaced with a more discrete version of the violin plot (see rightmost figure in \autoref{fig:teaser}). The question was framed as ``A random sample of 50 adults from Sweden were prescribed a new medication for one week. Another random sample of 50 adults from Sweden were assigned to a control group and given a placebo. Based on the information on the screen how confident are you that the medication decreases the body weight? Note the y-axis, higher values correspond to larger weight loss.''. The slider endpoints were labelled ``I have zero confidence in claiming an effect'', and ``I am fully confident that there is an effect.''.
	
	For this second experiment we used the same channels for sharing the link as in the first study and obtained 39 answers, of which two were discarded as they had not answered the background questions. Nine participants had expertise in ``Statistics and machine learning'', eight in ``VIS/HCI'', 14 in ``Social sciences and humanities'' and six in ``Physical and life sciences''.
	
	\subsection{Results}
	\subsubsection{Confidence Profiles and Cliff Effect}
	
	Table \ref{tab:desc2} shows the differences between subjective confidence when the underlying \pvalue\ was 0.06 versus 0.04.
	\begin{table}[t]
		\centering
		\caption{The sample mean, standard deviation, standard error of the mean, and the 2.5th and 97.5th percentiles of the difference in confidence when $p=0.04$ and $p=0.06$ in the second experiment.}
		\begin{tabular}{lrrrrr}
			\hline
			& Mean & SD & SE & 2.5\% & 97.5\% \\ 
			\hline
			Classic CI           & 0.07 & 0.12 &  0.02 & -0.22 & 0.28 \\ 
			Gradient CI          & 0.01 & 0.12 & 0.02 & -0.21 & 0.25 \\ 
			Continuous violin CI & 0.01 & 0.09 & 0.01 & -0.15 & 0.17 \\ 
			Discrete violin      & 0.06 & 0.18 &  0.03 & -0.17 & 0.50\\
			\hline
		\end{tabular}\label{tab:desc2}
	\end{table}
	The drop in confidence is again the largest with the classic CI with discrete violin CI having a similar drop. The relatively large standard error in the case of the discrete violin CI is explained by a small number of respondents that demonstrated a very large drop in confidence with the discrete violin CI. Overall the cliff effect seems to be much smaller than in the one-sample case (where the average drop was between 0.15--0.30, depending on the technique).
	
\begin{figure}[t]
		\centering
		\includegraphics[width=\columnwidth]{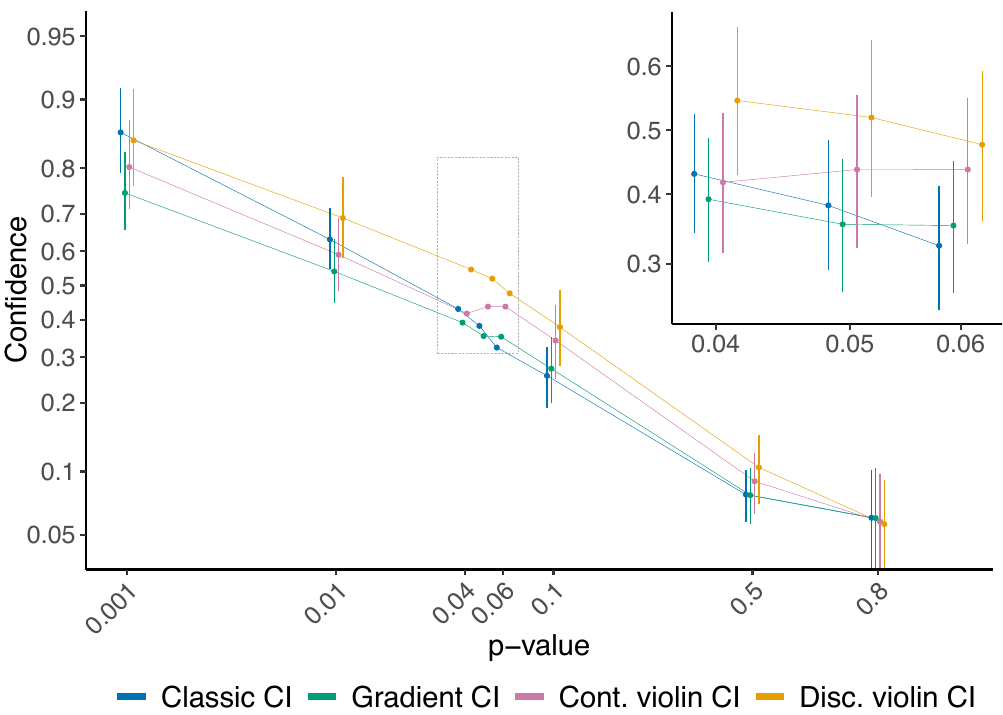}
		\caption{Posterior means of confidence and corresponding 95\% credible intervals for different visualization styles in the second experiment, on logit-logit-scale, with a zoom-in plot of the cases with $0.04 \leq p \leq 0.06$. A discontinuity in otherwise linear relationship between the true $p$-value and reported confidence indicates a cliff effect.}
		\label{fig:curves2}
	\end{figure}

\begin{figure}[t]
		\centering
		\includegraphics[width=\columnwidth]{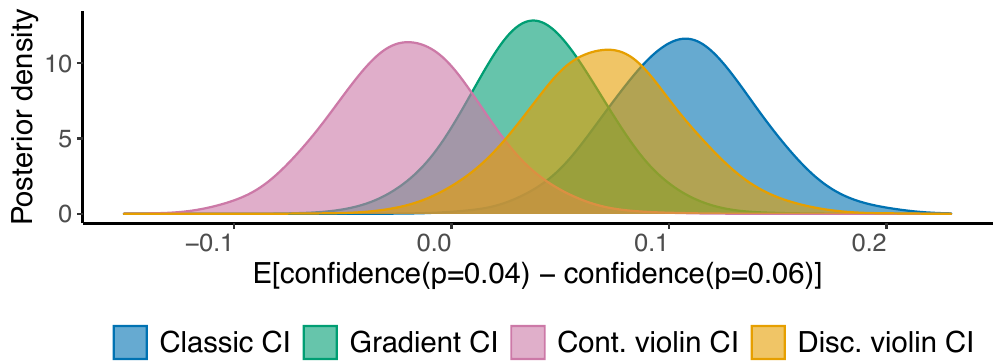}
		\caption{Posterior distributions of $\delta$, the drop in confidence around $p=0.05$, for different visualization styles in the second experiment.}
		\label{fig:cliff2}
	\end{figure}

	For analysing the results, we used the same multilevel model as for the first experiment. \autoref{fig:curves2} and \autoref{fig:cliff2} show the posterior mean curves of confidence and the posterior distributions of $\delta$ (the drop in confidence around 0.05). Compared with the first experiment, the overall confidence levels are smaller, for example with $p=0.04$ the average confidence is about 0.5 compared to 0.7 in the first experiment. There is a peculiar rise in the average confidence level for the continuous violin CI when the underlying $p$-value is 0.05 and 0.06 (although the credible intervals are wide)  
	but, overall, the differences between visualization styles are relatively small. Also, in contrast with the one-sample experiment, here we do not see clear signs of cliff effect or dichotomous thinking as the posterior mean curves are approximately linear (except, perhaps, for the classic CI where the posterior mean of $\delta$ is 0.1). As in the first experiment we saw no clear differences in the probability of an extreme answer between visualization styles.
	
	
	\subsubsection{Subjective Rankings}
	
\begin{figure}[h]
		\includegraphics[width=\columnwidth]{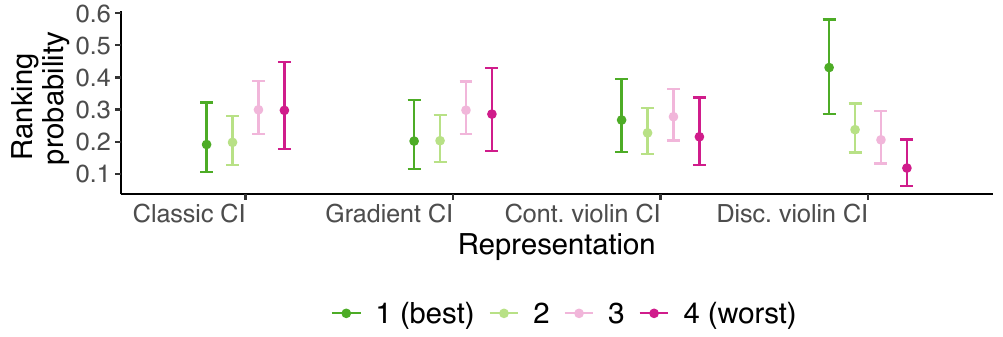}
		\caption{Subjective ranking probabilities and the corresponding 95\% credible intervals for visualization styles of the second experiment. A higher value for rank 1 indicates preference for the method while a higher value for rank 4 indicates distaste.}
		\label{fig:rankings2}
	\end{figure}

	As in the first experiment, we analysed the subjective rankings of the representation styles by Bayesian ordinal regression model where we explained the rank with visualization style and individual variance. \autoref{fig:rankings2} presents the ranking probabilities which indicate preferences towards the discrete violin CI plot (estimated to be the most preferred style by 42\% of the respondents). No clear differences emerge between other styles, and especially the classic CI and the gradient CI yield very similar results.
	
	\subsubsection{Qualitative Feedback}
	
	For this second controlled experiment, participants were also asked to comment on the limitations and benefits of each visualization. The fully categorized and raw data is, again, available in the supplementary material and we present the most frequent comments here. Classic CIs were reported as easy to read and analyse ($\times$ 12), space-efficient and a scalable visual representation ($\times$ 5), and as a standard visualization technique ($\times$ 5). Yet, some participants stated that they might call for dichotomous interpretations ($\times$ 5) and that they lack some information ($\times$ 12). 
	Continuous violin CI plots were said to provide more information than a classic CI ($\times$~2), but participants complained about the lack of explicit markers for the CI ($\times$ 6) and that the gradient was hard to see ($\times$ 3).
	Concerning discrete violin CI plots, participants noted that they are visually pleasing ($\times$ 2), that they provide more information than classic CIs ($\times$~10) and that seeing the discrete steps was very helpful---in comparison with the continuous violin plot ($\times$ 7). Still some participants highlighted that the gradient was hard to see ($\times$ 3) and that these plots could provide too much information in a single figure ($\times$ 2).
	Finally, gradient plots were deemed easy to interpret ($\times$ 8) but participants noted that the width was unnecessary ($\times$ 3), that some information was missing compared to gradient plots ($\times$ 4), and that the gradient was difficult to see ($\times$ 8).
	
	\section{Discussion}
	\label{sec:discussion}

	In line with previous findings~\cite{Lai:2010:DT,Hoekstra:2012:CIMD}, our results confirm that the classic CI visualization does not fix the cliff effect problem documented to be present in numerical and textual information. In fact, it appears that it may even \emph{increase} the cliff effect. At the same time, many participants preferred the graphical presentation of CIs over text, stating reasons such as the CI plot being clear and quick to grasp as well as scaling very well to multiple comparisons. 
	
	We found that more complex visualization styles reduced the cliff effect in the first one-sample experiment, and the violin CI plot, in particular, was also well received by the participants. 
	We found no clear differences between the interpretation of violin CI and gradient CI plots, which is in line with \cite{correll-gleicher-2014}. While we expected that these more novel visualization styles (violin and gradient CI plots) would introduce additional problems with interpretation due to unfamiliarity, their benefits seem to outweigh these negative effects. Some of the problems with violin CI plots could be explained by confusion with typical uses of a violin plot (as suggested by our feedback), namely as a method of visualizing observed data.
	
	The results from the second two-sample experiment suggest that the cliff effect might be a more common problem when comparing an estimate with a constant versus comparing two estimates, but further studies are needed to determine whether this is a general rule or just an artefact of our experimental setting or small sample size, especially as the lack of a clear cliff-effect in the two-sample experiment is in contradiction with the findings in~\cite{Belia:2005:RMC} that showed major problems in the interpretation of two-sample experiments (in a very different setting, however).
	
	Even though our convenience samples included researchers across a wide range of disciplines, it is unlikely to be fully representative of the general population of researchers using statistical analysis. Based on social media behaviour, survey feedback, and post-experiment discussions with some of the participants, our convenience sample likely contains disproportionate numbers of researchers with high knowledge and strong opinions on the topic of dichotomous thinking and the replication crisis. In particular, the links to the experiments were shared on the ``Transparent Statistics'' Slack channel which gathers HCI and VIS researchers who have argued for non-dichotomous interpretations of statistical results in their own work. We thus expect that our results likely downplay the average cliff effect compared with the much broader and heterogeneous scientific community. 
	
	Another factor which may have affected the answers of our participants is that we added explanatory texts to all the conditions to describe how they were created. This may have affected the responses of some participants, and it could be argued that the variation between the participants' answers and the observed cliff effect would have been greater without these explanations.
	As a third limitation, we observed a significant number of answers where the confidence increased with the underlying \pvalue. This was most clear in the VIS/HCI group with gradient and violin CI plots, and in general in the second experiment where the comparisons were more difficult. While these could explain some the estimated differences between representation styles, our sensitivity analyses, with samples where most of these counter-intuitive curves were removed, suggested only slight increases in the estimates of $\delta$ and identical general conclusions (see the supplementary material). As a further and more general limitation, we note that determining the ecological validity \cite{Quinan2015} of our experiment is, of course, non-trivial, e.g., in terms of whether one should differentiate dichotomous thinking and dichotomous graph reading.
	
	Despite the limitations, we expect that our results provide a valid lower estimate of the cliff effect in the broader scientific community and can be generalized into other statistics than just the sample mean. In general it is impossible to measure the potential costs of making dichotomous (and potentially wrong) interpretations \cite{Blakeley:2017:SSDE} as the costs are naturally context-specific. Nevertheless, given the negative effects of dichotomous thinking to the accumulation of scientific knowledge, we see violin and gradient CI plots as good alternatives for the classic CI as they significantly reduce the magnitude of the cliff effect. Given the already available tools for creating these types of visualizations, the long-term costs of adopting these new techniques are small and mainly related to increased space requirements.
	
	In contrast with most of the earlier studies on the cliff effect which have focused on psychologists or lay-people, we aimed to study the effect in a general population of researchers familiar with statistical methods. We used Bayesian modelling to take into account the individual-level variability in the answers and the uncertainty due to the parameter estimation leading to more realistic uncertainty assessments of our results than the traditional maximum likelihood estimation methods. We also provide a reproducible experiment with results available online and properly describe the questions we asked from the participants.
	
	\section{Conclusions and Future Work}
	\label{sec:conclusions}
	
	We provided analysis on the experiments on the cliff effect to study the effects of visual representation on interpreting statistical results. We found evidence that the problems with dichotomous thinking and the cliff effect are still common problems among researchers despite the amount of research and communications on this issue. In addition to educating researchers about this issue, we found that carefully chosen visualization styles can play an important role in reducing these phenomena.
	
	Our Bayesian multilevel model provides an illustration of how the data from relatively simple experiments can be analysed in a coherent modelling framework. It can give us more complex insights than simple descriptive statistics and avoids relying on the significance testing framework. The Bayesian approach also provides results that are easy to interpret, as everything is stated in terms of conditional probabilities which represent the state of knowledge. We hope this study encourages more model-based analysis in the VIS community in the future.
	
	All of our representations included a clear threshold for $p$-value 0.05 for comparative purposes. It would be interesting to study how visualization styles without this clear threshold would perform in similar settings. Also, quantile dot plots~\cite{Kay:2016:MBU, Fernandes:2018:UDU} (being discretized density plots) are similar to violin plots in terms of their information value but, as they lack some of the potential historical burden of more common violin plots, it would be interesting to compare the performance of these two representations in this setting.
	
	The consideration of space-efficient visual representations highlighted by some of our participants provides interesting avenues for future research. In line with recent work on interactive analyses and statistical visualization~\cite{Dragicevic:2019:EMA, gganimate, Hullman, Kale:2019:HOP}, we also anticipate that novel statistical representations free of the limitations of traditional printing constraints could have a positive impact both in general scientific communication and reducing dichotomous thinking. Indeed, our violin CIs could be made more space-efficient in order to better scale to multiple comparisons, for example by using interactive scaling. We therefore plan to study such solutions and their impact on statistical interpretations in future. As suggested by the discrepancy between the results of the first and second experiments, another avenue for further research is to study whether the cliff effect is stronger or more commonly occurring in settings where comparisons are made with respect to a constant reference point compared with multiple random variables.
	
	% use section* for acknowledgment
	\ifCLASSOPTIONcompsoc
	% The Computer Society usually uses the plural form
	\section*{Acknowledgments}
	\else
	% regular IEEE prefers the singular form
	\section*{Acknowledgment}
	\fi
	
	We thank P. Dragicevic, G. Cumming, and reviewers for their helpful comments. J. Helske was supported by Academy of Finland grants 311877 and 331817. S. Helske was supported by the Academy of Finland (331816, 320162) and the Swedish Research Council (445-2013-7681, 340-2013-5460).
	
	% Can use something like this to put references on a page
	% by themselves when using endfloat and the captionsoff option.
	\ifCLASSOPTIONcaptionsoff
	\newpage
	\fi

	% trigger a \newpage just before the given reference
	% number - used to balance the columns on the last page
	% adjust value as needed - may need to be readjusted if
	% the document is modified later
	%\IEEEtriggeratref{8}
	% The "triggered" command can be changed if desired:
	%\IEEEtriggercmd{\enlargethispage{-5in}}
	
	% references section
	
	% can use a bibliography generated by BibTeX as a .bbl file
	% BibTeX documentation can be easily obtained at:
	% http://mirror.ctan.org/biblio/bibtex/contrib/doc/
	% The IEEEtran BibTeX style support page is at:
	% http://www.michaelshell.org/tex/ieeetran/bibtex/
	\bibliographystyle{IEEEtran}
	\bibliography{statvis}

	% biography section
	% 
	% If you have an EPS/PDF photo (graphicx package needed) extra braces are
	% needed around the contents of the optional argument to biography to prevent
	% the LaTeX parser from getting confused when it sees the complicated
	% \includegraphics command within an optional argument. (You could create
	% your own custom macro containing the \includegraphics command to make things
	% simpler here.)
	%\begin{IEEEbiography}[{\includegraphics[width=1in,height=1.25in,clip,keepaspectratio]{mshell}}]{Michael Shell}
	% or if you just want to reserve a space for a photo:

	% if you will not have a photo at all:
	%\begin{IEEEbiographynophoto}{John Doe}
	%Biography text here.
	%\end{IEEEbiographynophoto}
	
	% insert where needed to balance the two columns on the last page with
	% biographies
	%\newpage
	\vskip -10pt
	\begin{IEEEbiography}[{\raisebox{5mm}{\includegraphics[width=1in,keepaspectratio]{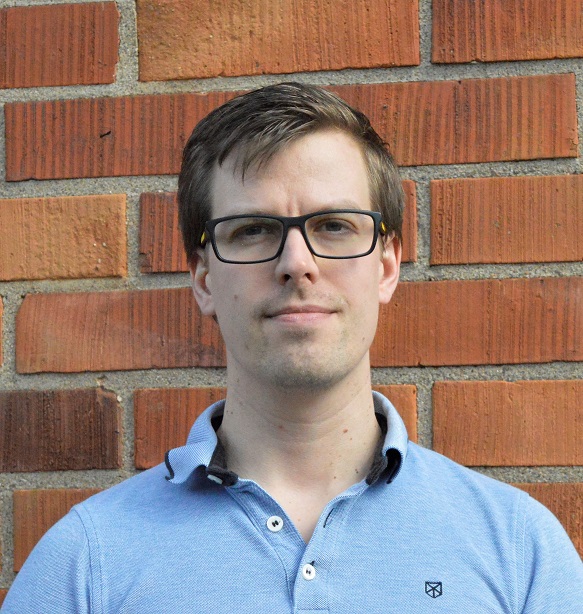}}}]{Jouni Helske} is a senior researcher at the University of Jyv\"askyl\"a, Finland, from where he received the Ph.D. degree in statistics. He was previously a postdoctoral researcher at the Link\"oping University, Sweden. His research has focused on state space models, Markov chain Monte Carlo and sequential Monte Carlo methods, information visualization, and his current research focuses on Bayesian causal inference.
	\end{IEEEbiography}
	
	\begin{IEEEbiography}[{\raisebox{5mm}{\includegraphics[width=1in,keepaspectratio]{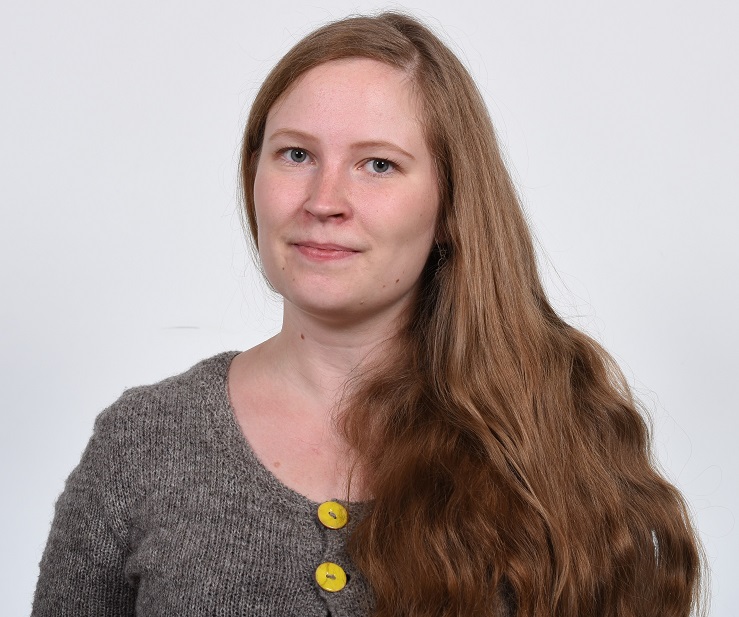}}}]{Satu Helske} is a senior researcher in sociology at the University of Turku, Finland. She received the Ph.D. degree in statistics from the University of Jyv\"askyl\"a, Finland, after which she worked as a postdoctoral researcher at the University of Oxford, UK, and at Link\"oping University, Sweden. She works at the crossroads of sociology and statistics, with her main focus being on longitudinal and life course analysis.
	\end{IEEEbiography}
	
	\begin{IEEEbiography}[{\raisebox{5mm}{\includegraphics[width=1in,keepaspectratio]{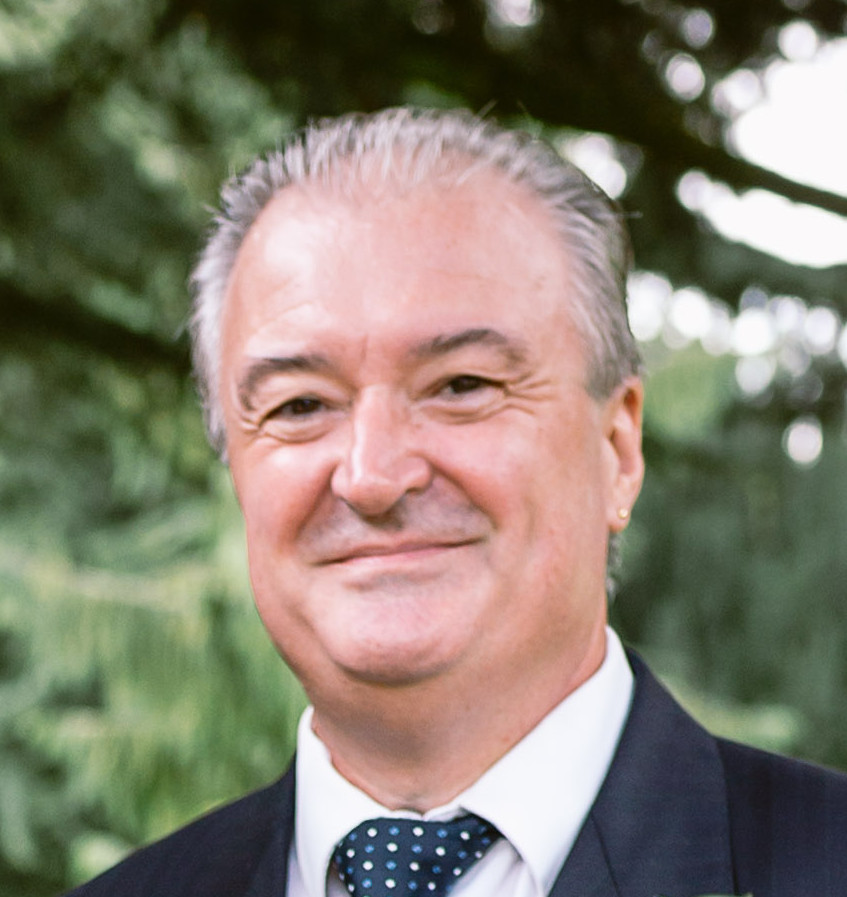}}}]{Matthew Cooper} is senior lecturer in information visualization with the University of Link\"{o}ping. He was awarded a PhD in Chemistry by the University of Manchester, UK, in 1990 but joined the Manchester Visualization Centre in 1996. He joined the University of Link\"{o}ping in 2001. His current interests lie in visual representations and analytical methods for multivariate and temporal data, and the user-centred evaluation of visualization techniques.
	\end{IEEEbiography}
	
	\begin{IEEEbiography}[{\raisebox{5mm}{\includegraphics[width=1in,keepaspectratio]{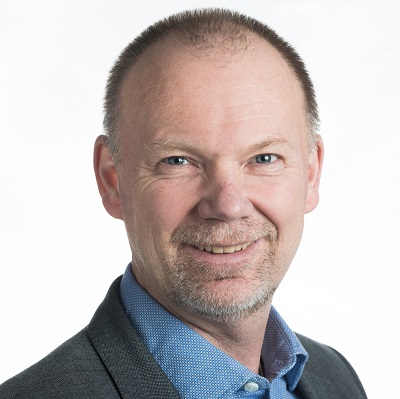}}}]{Anders Ynnerman} is a Professor in scientific visualization at Link\"{o}ping University and is the director of the Norrk\"{o}ping Visualization Center C. His research has focused on interactive techniques for large scientific data in a range of application areas. 
		%He received the Ph.D. degree in physics from Gothenburg University, Sweden in 1992. 
		He is a member of the Swedish Royal Academy of Engineering Sciences and the  Royal Swedish Academy of Sciences and in 2018 he received the IEEE VGTC technical achievement award.
	\end{IEEEbiography}

	\begin{IEEEbiography}[{\raisebox{5mm}{\includegraphics[width=1in,keepaspectratio]{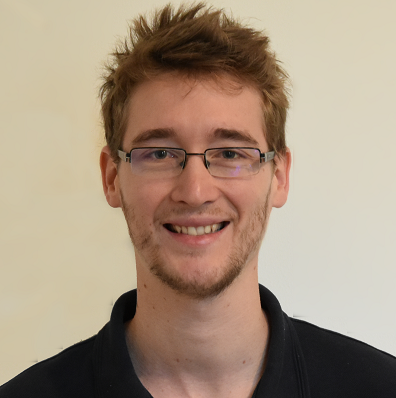}}}]{Lonni Besan\c{c}on} is a postdoctoral fellow at Link\"oping University, Sweden. He received the Ph.D. degree in computer science at University Paris Saclay, France. He is particularly interested in interactive visualization techniques for 3D spatial data relying on new input paradigms. Recent work focuses on the visualization and understanding of uncertainty in empirical results in computer science.
	\end{IEEEbiography}
	
	% You can push biographies down or up by placing
	% a \vfill before or after them. The appropriate
	% use of \vfill depends on what kind of text is
	% on the last page and whether or not the columns
	% are being equalized.
	
	%\vfill
	
	% Can be used to pull up biographies so that the bottom of the last one
	% is flush with the other column.
	%\enlargethispage{-5in}

	% that's all folks
\end{document}